# gTASEP with attraction interaction on lattices with open boundaries

N. C. Pesheva,  N. Zh. Bunzarova,

*Institute of Mechanics, Bulgarian Academy of Sciences, 1113 Sofia, Bulgaria*

**Abstract**

We study a model of aggregation and fragmentation of clusters of particles on an open segment of a single-lane road. The particles and clusters obey the stochastic discrete-time discrete-space kinetics of the Totally Asymmetric Simple Exclusion Process (TASEP) with backward ordered sequential update (dynamics), endowed with two hopping probabilities, p and $p_m$. The second modified probability, $p_m$, models a special kinematic interaction between the particles belonging to the same cluster. This modification is called generalized TASEP (gTASEP) since it contains as special cases TASEP with parallel update and TASEP with backward ordered sequential update for specific values of the second hopping probability $p_m$. We focus here on exemplifying the effect of the additional attraction interaction on the system properties in the non-equilibrium steady state. We estimate various physical quantities (bulk density, density distribution, and the current) in the system and how they change with the increase of $p_m$ ($p < p_m < 1$). Within a random walk theory we consider the evolution of the gaps under different boundary conditions and present space-time plots generated by MC simulations, illustrating the applicability of the random walk theory for the study of gTASEP.

## 1. Introduction

The generalized TASEP was first proposed in [1] and later studied on a ring in [2-4] as an integrable generalization of the TASEP with an additional interaction, which favors (when $p < p_m$) the clustering of particles. The model incorporates two extreme cases: the TASEP with parallel update (PU) when $p_m=0$ is set (see, e.g., Refs [5,6]) and the case with all particles irreversibly merging (when $p_m=1$) into a single cluster moving as an isolated particle. The latter case is that of the irreversible aggregation (IA), studied in Refs [7-9]. The gTASEP reduces to the extensively studied ordinary TASEP with backward ordered sequential update (BOSU) when $p_m=p$ (see, e.g., Refs [10,11]).

TASEP is one of the most studied models of non-equilibrium phenomena and has attracted the interest of both mathematicians and physicists Refs [12-13]. It is one of the simplest exactly solved models of driven many-particle systems, with bulk particle conserving stochastic dynamics [14-15]. In Nature true equilibrium phenomena are rarely found, they are rather an idealization of real processes. Most of the phenomena are non-equilibrium and most of the systems (in this number the living systems) are exchanging matter and/or energy with their surroundings, sustaining non-trivial currents. However, today's understanding of non-equilibrium systems lags behind the knowledge and understanding of equilibrium systems. The importance of developing a fundamental and comprehensive theory of systems far from equilibrium was recognized quite a time ago and since then very active research is under way of various non-equilibrium systems [16-18].

TASEP has found a number of applications to biological transport [13,19], vehicular traffic flow [20,21], forced motion of colloids in narrow channels [22], transport of data packets on the internet, just to mention a few.

Aggregation and fragmentation of clusters of arbitrary size arise in many physical-chemical processes: aerosol physics, polymer growth, aggregation of platelets, protein aggregation and even in astrophysics. In medicine, the ability to control protein aggregation could be a useful tool for developing new drug. Irreversible aggregation may have quite a destructive role, e.g., many neurodegenerative diseases (Alzheimer's disease, Parkinson's disease, and prion diseases) are characterized by intracellular aggregation and deposition of pathogenic proteins [23].

The gTASEP model we study here is designed to be simple rather than realistic; nevertheless, it still can be helpful in understanding various types of systems in Nature and in particular in biophysics.

The paper is organized as follows. In Section 2 we formulate the model with open boundary conditions and present the phase diagrams obtained in previous studies [7-9]. In Section 3 we present results of MC simulations of gTASEP with open boundaries illustrating the effect of the attraction interaction on some system properties. Within a random walk theory we consider the evolution of the gaps under different boundary conditions. Short overview of the main results and some outlooks for further studies are given in the Conclusion.

2. **The model**

As mentioned in the Introduction the dynamics of the gTASEP is based on the standard TASEP with BOSU, however, there is a second modified hopping probability $p_m$ in addition to the standard hopping probability p (see Fig. 1 where schematic image of gTASEP is presented). Here we consider the gTASEP with open boundary conditions, i.e., we consider an open one-dimensional lattice of $L$ sites. Every site is either empty $\tau_i = 0$ or occupied $\tau_i = 1$, where $\tau_i$ is an occupation number associated with a site $i$ ($1 < i < L$). Each configuration $\{\tau_i\}, i = 1, \cdots, L$ update of the system starts with the update of the last site of the chain $L$. If site $i = L$ is occupied, the particle at it leaves the system with probability $\beta$ and stays in place with probability $1 - \beta$.

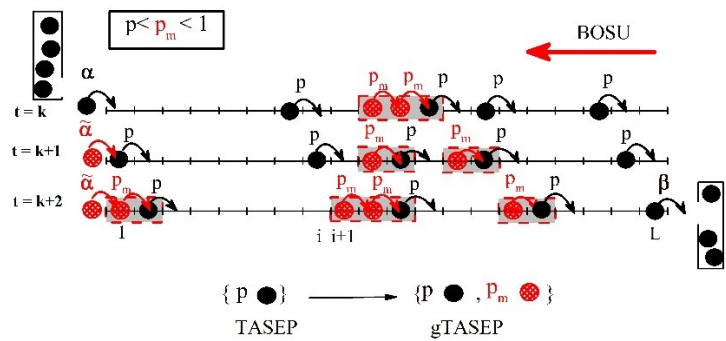

**Fig. 1.** The generalized TASEP with open boundaries – schematic image. Time evolution of the system configurations is presented at three consecutive time steps to illustrate the update rules in the case $p < p_m < 1$ (attraction interaction).

Isolated particles and the first particle of a cluster on the right may move one site to the right with probability p (or to stay at place with 1-p). Particles, which belong to a cluster (except the head particle) may move one site to the right, provided the particle in front of them has moved at the same time step, with a modified probability $p_m$ or stay immobile with probability $1 - p_m$. When $p_m > p$ – "attraction" interaction between particles in a cluster is modelled. The left boundary condition is also modified accordingly (it was suggested in [24] and independently in [7]) to ensure consistency with the update rules in the bulk. If site i = 1 was empty at the beginning of the current update, a particle enters the system with probability α, or site i = 1 remains empty with probability $1 − α$. If site i = 1 was occupied at the beginning of the current moment of time, but became empty under its current update, then $\hat{α} = \min\{α\, p_m/p, 1\}$.

**Phase diagrams of gTASEP when $p_m$=1 and $p_m$=0.99**

Our extensive Monte Carlo simulations of the model system, obeying the above generalized TASEP dynamics in the case of irreversible aggregation [7,8] point out to a phase diagram in the (α,β) plane containing three phases with novel topology. As shown in Fig. 2 the phase diagrams of the stationary phases of the gTASEP in the particular cases $p_m = 1$ (IA) and $p_m > p$ ("attraction" interaction) are different. The model with IA has three stationary phases: (1) a many-particle one, MP consisting of two sub-regions MP I and MP II; (2) a phase with completely filled configuration (CF); and (3) a mixed (boundary perturbed) MP+CF phase, shown in Fig. 2(a). The regions MPI and MPII represent a phase, containing a macroscopic number of particles or clusters. The bulk density can take any value from zero to one. The local density profile is flat up to the first site, $ρ_b = α/p \equiv \hat{α}$, but the two regions differ by its shape near the chain end. Numerically we find $ρ_L = α/β$. The phase CF represents a chain completely filled with particles with a stationary current of particles $J = β$. The phase MP+CF is a combination of many particle configurations and nonzero probability of complete filling of the chain P(1) in the finite-size limit. The chain is completely filled at the bulk and up to the last site, $\rho_b = \rho_L = 1$. We have shown that the density at the first site is $\rho_1^{MP+CF} = 1 - (1/\alpha - 1/p)\beta$.

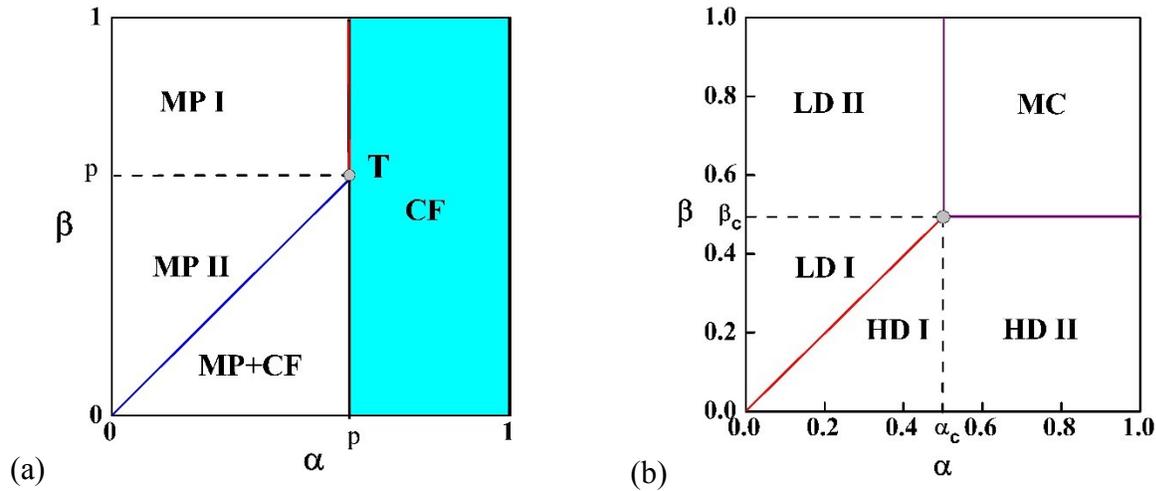

**Fig. 2.** Phase diagrams in the plane of injection/ejection rate probabilities (i.e., in the (α,β)-plane) of: **(a)** the gTASEP with p = 0:6 and $p_m$ =1 (IA). The critical injection/ejection rates are estimated $α_c$ (0.6; 1) = $β_c$ (0.6; 1) = p; and **(b)** the gTASEP with $p_m$ = 0.99, where aggregation-fragmentation of clusters occurs. The critical injection/ejection rates are $α_c$ (0.6; 0.99) = $β_c$ (0.6; 0.99) = 0.502. For more details see text.

The unusual phase transition, characterized by jumps both in the bulk density $\rho_b(\alpha)$ and the current $J(\alpha)$, found in [7], takes place across the boundary $\alpha = p$ between the MPI and CF phases. The topology of the phase diagram in the case of IA (when $p_m = 1$) (Fig. 2(a)), changes sharply to the one, corresponding to the ordinary TASEP, however, with ($p$, $p_m$)-dependent triple point ($\alpha_c, \beta_c$), as soon as the modified probability $p_m$ becomes less than unity and aggregation-fragmentation of clusters is allowed (see Fig. 2(b)) [9]. For example we have estimated $\alpha_c$ (0.6; 0.99) = $\beta_c$ (0.6; 0.99) = 0.502 as compared to $\alpha_c$ (0.6; 0.6) = $\beta_c$ (0.6; 0.6) = 0.3675, which are the critical values for the ordinary TASEP.

3. Results

**Simulation results**

The "interactions" among the particles determine the way the particles influence each others' movement. The case when $p_m > p$ is termed as "attraction" interaction, since the particles in a cluster have higher probability to move than the head particle and thus they have higher chance to stay together in the cluster than to split. The effect of the additional interaction between the particles can be observed in Figs 3, 4 below. In Fig. 3 one can observe how the density profile $\rho(x)$ ($x=i/L$, $x \in [0,1]$) at one point in the phase space, i.e., at $\alpha = 0.5$, $\beta = 0.5$, in the gTASEP is changed as the second modified probability is increased from $p_m = p$ (TASEP with BOSU) to $p_m = 1$ – the case, when IA of clusters and particles occurs. At $p_m = 0.6$ and $p_m = 0.9$ the density profile is that of a system in MC phase. The density profile changes continuously to the density profile, found at $p_m = 0.99$, which is a density profile on the coexistence line (between the LDI and HDI phases), practically at the triple point, since for $p_m=0.99$ one has $\alpha_c = \beta_c = 0.502$ (see Fig 2(b)). The density profile at $p_m =1$ is a profile on the coexistence line between the MPII and MP+CF phases of the phase diagram in the case of IA (see Fig 2(a)).

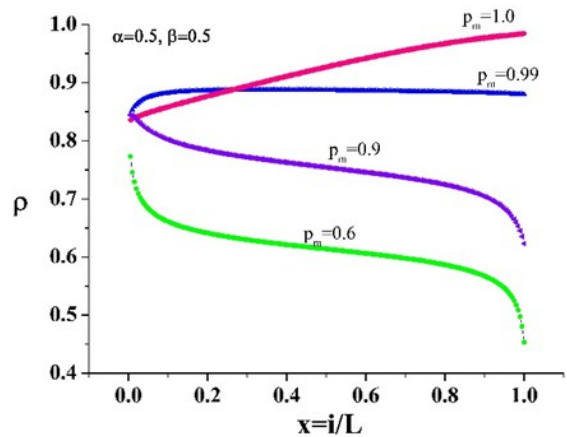

**Fig. 3.** The effect of the additional interaction ($p < p_m < 1$) between the particles on the density profiles at the phase point $\alpha=0.5$, $\beta=0.5$ is illustrated in the gTASEP for different values of $p_m$: $p_m =0.6$, $p_m =0.99$, and $p_m =1.0$ (IA).

In terms of real traffic, the case $p_m > p$ models the natural tendency of a driver to catch up with the car ahead. Thus clusters of synchronously moving particles or cars may appear, leading to higher throughput (current) in the system at any density – see Fig. 4(a) where the current and the bulk density in the system (at α = 0.5, β = 0.5) and (b) where the fundamental diagram (the current versus bulk density) for different values of $p_m$ are shown below.

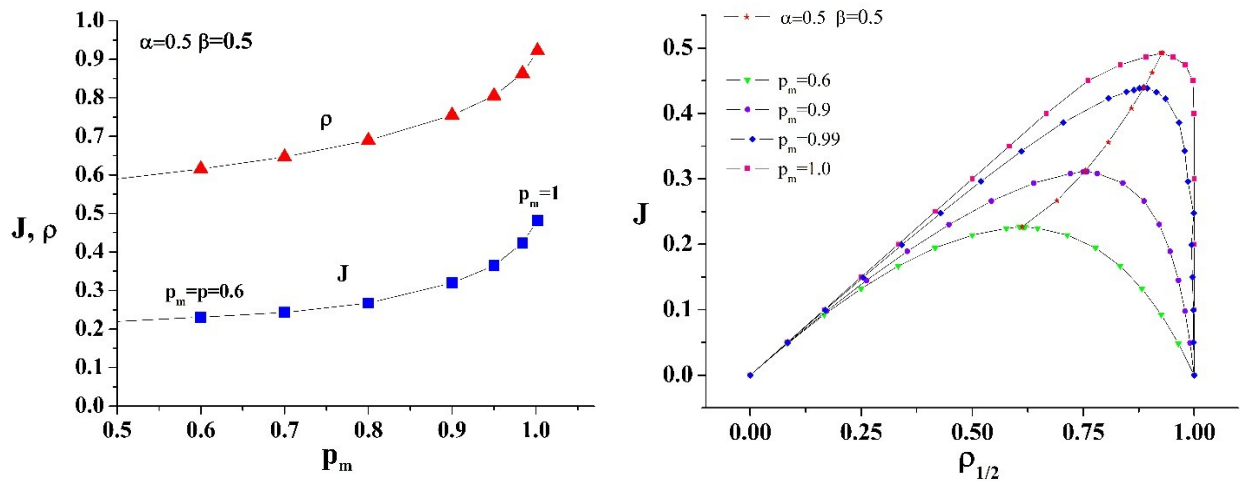

**Fig. 4. (a)** The current and the bulk density in the system at α=0.5, β=0.5 as functions of the modified probability $p_m$; **(b)** The fundamental diagram (current J versus bulk density ρ) at different values of $p_m$: $p_m$=0.6 (TASEP with BOSU) -- green symbols-solid line; gTASEP with $p_m$=0.99 -- blue symbols-solid lien, and gTASEP with $p_m$=1.0 (IA) – pink symbols-solid line. The red solid stars show the result for α=0.5, β =0.5 and different values of $p_m$.

The density at the middle of the chain, $\rho_{1/2}$ at x=1/2, (where x=i/L) is a very good approximation of the bulk density ρ=$\rho_b$ in the system and we use it here instead of ρ. The fundamental diagram of gTASEP was obtained first by Hrabak [17] for gTASEP with periodic BC. Here we present our simulation results obtained for gTASEP with open boundary conditions for a system of L=200 sites. One can see in Fig. 4(b) that the current J at fixed ρ is increasing with $p_m$. The maximal value of the current at fixed $p_m$ (shown by the (red) stars) is shifting with $p_m$ towards higher densities – a very useful and desirable property for real traffic. One can see in Fig. 1(a) that at $p_m$=1 the current is more than twice the value at $p_m$=0.6 and the bulk density is more than 0.9.

**Random walk theory in the generic case of attraction ($p_m$ < 1)**

One of the theoretical methods, developed in Refs. [7-9], which we concisely present and use here, is based on the study of the time evolution of single gaps in different regions of the CF phase. In this situation, one needs to use the dual representation of the system configurations, i.e., the empty sites positions, instead of particle positions. Such a representation illustrates a specific dynamics of the inter-cluster gaps resulting in the property that the width of each gap performs a random walk.

The problem is rather complicated since when when $p_m < 1$ the probability of appearance of a gap is position dependent. In contrast to the case of $p_m = 1$, here we show that when $\beta \neq p$, the gap width performs a special, position dependent random walk. We start by finding out the probability of a single gap appearance under boundary conditions corresponding to the CF phase and consider the first step in the time evolution of the gap width. Having in mind the update rules of the system one can see that the gap appears at sites $2 \leq i \leq L$ as a result of the left boundary condition. Thus the gap width increases by one site with probability $p_g(i) = (1-p)p_m^{L-i-1}\beta$, decreases by one site with probability $q_g(i) = [(1-\beta) + (1-p_m^{L-i-1})\beta]p = (1 - p_m^{L-i-1}\beta)p$, and remains the same with probability $r_g(i) = 1 - p + p_m^{L-i-1}\beta(2p-1)$. We average the gap width evolution over the initial probabilities given by $P_{L-k}(p,p_m) = (1-p_m)p_m^k\beta$, $k = 0,1,...L-2$, and $P_1(p,p_m) = (1-\tilde{\alpha})p_m^{L-1}\beta$. One can conclude that on the average a single-site gap will grow after the first time step of its evolution when

$$\beta > p\frac{p_m(1+p_m)}{1+p_m^{L-1}}$$

Thus when $p_m \to 1$ and $L$ is fixed, or $L \to \infty$ so that $p_m^L \to 1$ this condition simplifies to $\beta > p$. However, for fixed values of $p_m$ close to 1, $p_m^L$ will decrease to zero as $L \to \infty$. On the grounds of our random walk theory and the computer simulations, we conjecture that the simple criteria $\beta > p$ for growing gaps, and $\beta < p$ for decreasing gaps, hold true. Having that in mind we infer that in the upper region $(p < \alpha \leq 1] \times (p < \beta \leq 1]$ of the CF phase (see Fig 2(a)) a maximum-current phase (see Fig 2(b)) will appear. Its local density profile satisfies the inequalities $\rho_1 = 1 > \rho_{l/2} > \rho_L$, which follow from the conditions $\hat{\alpha} = 1$, and the larger probability of gap formation near the end of the chain. In the lower region $(p < \alpha \leq 1] \times (0 < \beta < p]$ of the CF phase the gaps are scarce, small and short-living, which is characteristic of a high-density phase. The left-hand side of the local density profile bends upward to $\rho_1 = 1$.

We can find additional information in the different gaps evolution regimes in regions LDI (shown in Fig. 5(a) an Fig. 6(a) below) and HDI (Fig. 5(b) an Fig. 6(b)). In both cases $\alpha < p$, which implies $\hat{\alpha} < 1$, and gaps can appear at the first site $i = 1$ and evolve throughout the chain. In the LDI region the gaps are wide and long living, while in the HDI region (Fig. 6(b)) they are small, scarce and very short living. These features may explain the large difference in the particle densities in the low-density and high-density phases.

For easier comparison and illustration of the different gaps evolution regimes we present the space-time plots, generated by MC simulations, for TASEP with $p_m$=0.6, gTASEP with $p_m$=0.99 and gTASEP with $p_m$=1 (IA), at the same points in the phase space, i.e., at $\alpha$ =0.2, $\beta$ =0.7; $\alpha$ =0.31, $\beta$ =0.3, and at $\alpha$=0.7, $\beta$=0.7. These points belong to different phases of gTASEP when $p_m$ is varied These different phases are indicated in the figure captions below every space-time plot.

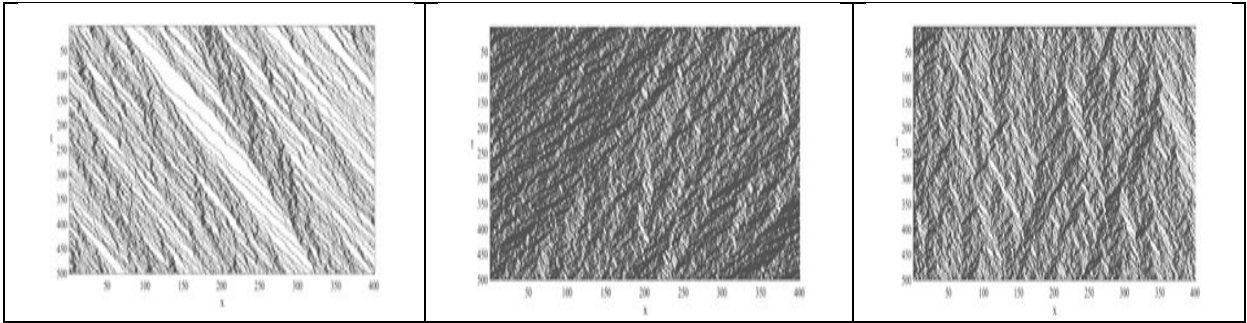

**Fig. 5.** Space-time plots (time is flowing downward in the vertical direction) for a system of L = 400 sites at representative points in the phase space in the case of TASEP with BOSU ($p_m$=p=0.6) : **(a)** α =0.2 β =0.7 (LDI) **(b)** α =0.31 β =0.3 (HD) and **(c)** α=0.7 β=0.7 (MC).

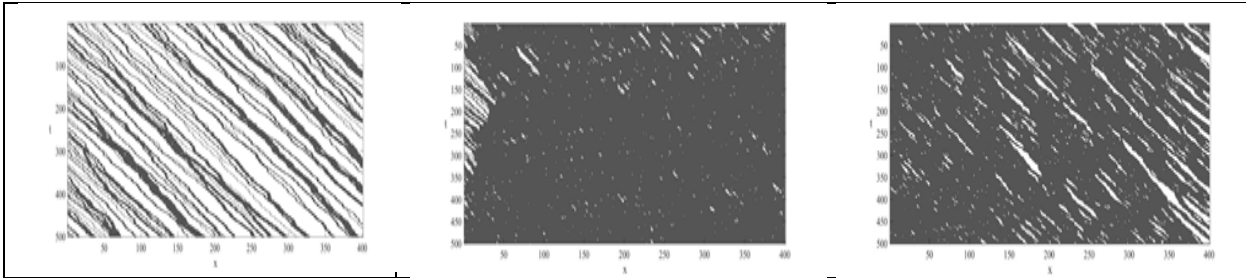

**Fig. 6.** Space-time plots of the gTASEP with p = 0.6, $p_m$ = 0.99, showing the gaps evolution in: **(a)** phase LDI (α = 0.2, β = 0.7 **(b)** phase HD (α =0.31 β =0.3) and **(c)** phase MC (α=0.7 β=0.7).

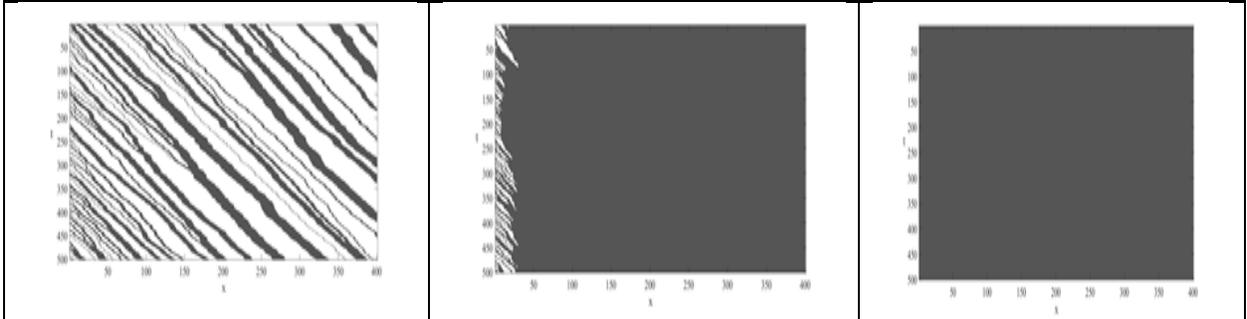

**Fig. 7.** Space-time plots of the gTASEP with p=0.6 and $p_m$=1 (IA) showing the gaps evolution in: **(a)** phase MPI (α = 0.2, β = 0.7; **(b)** phase MPII ( α =0.31, β =0.3); and **(c)** phase CF (α=0.7 β=0.7).

## 4. Conclusion

The gTASEP with attraction interaction ($p_m$ > p) was studied on open lattice segments. We obtain various physical quantities in the system by means of MC simulations (bulk density, density distribution, and the current) and how they change when increasing the probability $p_m$ (p < $p_m$ ). We find that the main effect of increasing the modified hopping probability $p_m$ is increase in the bulk density, and the current (Figs 3,4) in the system. The topology of the phase diagram remains the same as for the ordinary TASEP, however, the critical injection/ejection probabilities increase (at fixed p) with $p_m$ . Further studies are needed to get a more detailed understanding of the system behavior, like finding the cluster distribution in different phases, fluctuation properties of the total number of particles, etc.

**Acknowledgements**

The authors gratefully acknowledge fruitful discussions with their late colleagues and coauthors, Professor V. B. Priezzhev and Professor J. G. Brankov, held at earlier stage of the present study.

Partial financial support of the Bulgarian MES by Grant No. D01-221/03.12.2018 for NCDSC—part of the Bulgarian National Roadmap on RIs is also thankfully acknowledged.